\documentclass[10pt,journal,compsoc]{IEEEtran}
\usepackage{CJK}
\usepackage{indentfirst}
\usepackage{multirow}
\usepackage{amsmath}
\usepackage{cases}
\usepackage{tabularx}
\usepackage{tikz}  
\usetikzlibrary{arrows,graphs} 
\usepackage{mathpazo}

\tikzset{elegant/.style={smooth,thick,samples=50,cyan}}
\tikzset{eaxis/.style={->,>=stealth}}

\usepackage{algorithm}  
\usepackage{algpseudocode}  
\ifCLASSOPTIONcompsoc
\usepackage[nocompress]{cite}
\else
\usepackage{cite}
\fi
\begin{document}
	%
	\title{The Application of Bipartite Matching in Assignment Problem}
	%
	%
	%
	
	\author{Feiyang Chen, Nan Chen , Hanyang Mao, Hanlin Hu}

	%
	%

	\markboth{Chuangxinban Journal of computing, May~2018}%
	{Shell \MakeLowercase{\textit{et al.}}: }

	\IEEEtitleabstractindextext{%
		\begin{abstract}
			The optimized assignment of staff is of great significance for improving the production efficiency of the society. For specific tasks, the key to optimizing staffing is personnel scheduling. The assignment problem is classical in the personnel scheduling. In this paper, we abstract it as an optimal matching model of a bipartite graph, and propose the Ultimate Hungarian Algorithm(UHA). By introducing feasible labels, iteratively searching for the augmenting path to get the optimal match(maximum-weight matching). And we compare the algorithm with the traditional brute force method, then conclude that our algorithm has lower time complexity and can solve the problems of maximum-weight matching more effectively.
		\end{abstract}}
		
	
	\maketitle

	\IEEEdisplaynontitleabstractindextext

	%
	\IEEEpeerreviewmaketitle

	\IEEEraisesectionheading{\section{Introduction}\label{sec:introduction}}

	%
	%
	%
	%
	
	\IEEEPARstart{M}{otivated} by the increasing in labor costs, employing multi-skilled workers has become an important way to raise the personnel utilization. For a specific task, personnel scheduling is the key to optimizing staffing. Assignment problem is one of the classic problem in the staff scheduling. $m$ workers match $n$ tasks, and each worker has different efficiency in each task, determine a task allocation plan which makes the highest gross efficiency. Traditional approach to this problem is manually allocate the tasks according to the leader's understanding of the workers, which is time-consuming, laborious, inefficient and relying on experience. However, this approach can not meet the requirements of quantitative, rapid and automatic scientific management. Therefore, it is necessary to introduce the theories of operational research to solve the assignment problem with computer assistance.
	\par
	In this paper, we abstract the maximum-weight matching model from assignment problem, and propose UHS to solve it. The main idea of UHS is to figure out the maximum-weight matching by searching augmenting path. Compared to the traditional method, our new method liberates the human labor and find maximum-weight matching more precisely.

	\section{Preliminary}
	\subsection{Assumption}
	When dealing with assignment problem, we use subsets of vertices to stand for workers and tasks, and edges between two vertex from different subsets means the worker and task can be matched. We assume that each worker's ability for tasks is quantifiable, which reflects in the weights of edges. Meanwhile, each worker can match up with only one task in our assumption.
	\subsection{Definition}
	\subsubsection{Bipartite graph}
	Given a unidirectional graph $G = (V,E)$, if the vertex set can be partitioned into $V = P \cup Q$,  $P$ and $Q$ are disjoint and all edges in $E$ go between $P$ and $Q$, then $G$ called a Bipartite graph.
	\subsubsection{Matching}
	A Matching is a subset of edges $M \subseteq E$ such that for all vertices $v \in V$, at most one edge of $M$ is incident on $v$. A maximum matching is a matching $M$ for any matching $M'$, we have $|M'| \subseteq |M|$. And maximum-weight matching is the matching of the largest sum of weights.
	\subsubsection{Augmenting path}
	A path is augmenting  for a matching $M$  if it alternates between edges in the matching and edges not in the matching, and the first and last vertices are unmatched.
	\section{Design of the UHA}
	\subsection{Feasible Labeling}
	The majority of realistic matching problems are much more complex. This added complexity often stems from graph labeling, where edges or vertices labeled with quantitative attributes, such as weights, costs, preferences or any other specifications, which adds constraints to potential matches.
	
	A common characteristic investigated within a labeled graph is a known as feasible labeling, where the label, or weight assigned to an edge, never surpasses in value to the addition of respective vertices' weights. This property can be thought of as the triangle inequality.
	\begin{equation}
	l(x) + l(y) \geq w(x,y)\ \forall x \in X, y \in Y
	\end{equation}
	Where $X$ is the set of nodes on one side of the bipartite graph, $Y$ is the other set of nodes, $l(x)$ is the label of $x$, etc., and $w(x,y)$ is the weight of the edge between $x$ and $y$.
	
	\begin{figure}[ht]
		
		\centering
		\includegraphics[scale=0.5]{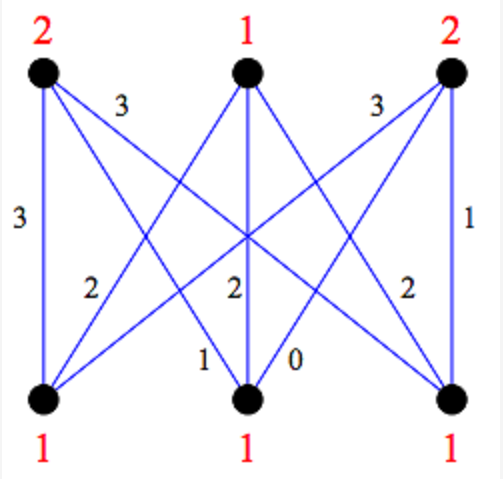}
		
		\centering\caption{A simple feasible labeling is just to label a node with the number of the largest weight from an edge going into the node, where labels are in red. }
		\label{fig:label}
	\end{figure}
	\subsection{Proof}
	
	\textbf{Lemma: A feasible labeling on a perfect match returns a maximum-weighted matching.}
	\par
	Suppose each edge $e$ in the graph $G$ connects two vertices, and every vertex is covered exactly once. With this, we have the following inequality:
	\begin{equation}
	\sum_{e\ \epsilon\ E} w(e) \leq \sum_{e\ \epsilon\ E } \big(l(e_x) + l(e_y)\big) = \sum_{v\ \epsilon\ V} l(v)
	\end{equation}
	Where $M'$ is any perfect matching in $G$ created by a random assignment of vertices, and $l(x)$ is a numeric label to node $x$.
	This means that is an upper bound on the cost of any perfect matching.
	
	Now let $M$ be a perfect match in $G$, then 
	\begin{equation}
	w(M) = \sum_{e\ \epsilon\ E} w(e) = \sum_{v\ \epsilon\ V}\ l(v)
	\end{equation}
	So \(w(M') \leq w(M)\) and $M$ is optimal. 
	\subsection{Implementation}
	The algorithm starts by labeling all nodes on one side of the graph with the maximum weight. This can be done by finding the maximum-weighted edge and labeling the adjacent node with it. Additionally, match the graph with those edges. If a node has two maximum edges, do not connect them.
	\par
	If the matching is perfect, the algorithm is done as there is a perfect matching of maximum weights. Otherwise, there will be two nodes that are not connected to any other node. If this is the case, begin iterating.
	\par
	Improve the labeling by finding the non-zero label vertex without a match, and try to find the best assignment for it. Formally, the UHA matching algorithm can be executed as defined in the Algorithm 1.
	\par 
	\begin{algorithm}[htb]  
		\caption{UHA}  
		\label{alg:Framwork}  
		\begin{algorithmic}[1]
			\Require  
			The labeling $l$, an equality graph $G_{l}=(V,E_{l})$, an initial matching $M$ in $G_{l}$, and an unmatched vertex $v\in V$ and $u\notin M$;
			\Ensure  
			Maximum-weighted matching;  
			\State We will keep track of a candidate augmenting path starting at the vertex $u$;  
			\label{code:fram:extract}  
			\State If the algorithm finds an unmatched vertex $v$, add on to the existing augmenting path $p$ by adding the $u$ to $v$ segment;  
			\label{code:fram:trainbase}  
			\State Flip the matching by replacing the edges in $M$ with the edges in the augmenting path that are not in $M$ (in other words, the edges in $E_{l} - M$);  
			\label{code:fram:add}  
			\State  $S \subseteq X$ and $T \subseteq Y$ where $S$ and $T$ represent the candidate augmenting alternating path between the matching and the edges not in the matching;  
			\label{code:fram:classify}  
			\State Let $N_{l}(S)$ be the neighbors to each node that is in $S$ along edges in $E_{l}$ such that \(N_l(S) = \{v|\forall u \in S: (u,v) \in E_l\}\);
			\State If \(N_l(S) = T\), then we cannot increase the size of the alternating path (and therefore can not further augment), so we need to improve the labeling;
			\State Let $N_{l}(S)$ be the neighbors to each node that is in $S$ along edges in $E_{l}$ such that \(N_l(S) = \{v|\forall u \in S: (u,v) \in E_l\}\);
			\State Let $\delta_{l}$ be the minimum of \(l(u) + l(v) - w(u,v)\) over all of the $ u \in S$ and \(v \notin T\);
			\State Improve the labeling $l$ to $l'$:
			\par
			If $r \in S$, then $l'(r) = l(r) - \delta_{l}$;
			\par
			If $r \in T$, then $l'(r) = l(r) + \delta_{l}$;
			\par
			If $r \notin S$ and $r \notin T$, then $l'(r) = l(r)$;
			
		\end{algorithmic}  
	\end{algorithm}   
\par
Each step will increase the size of the matching  $M$ or it will increase the size of the set of labeled edges, $E_{l}$. This means that the process will eventually terminate since there are only a limited number of edges in the graph $G$.
\par
When the process terminates,  $M$ will be a perfect matching. By the Kuhn-Munkres theorem, this means that the matching is a maximum-weight matching.
\section{Analysis}
At the first step, the algorithm adds one edge to the matching and this happens $O(|V|)$ times. Then It takes  $O(|V|)$ time to find the right vertex for the augmenting (if there is one at all), and it is $O(|V|)$ time to flip the matching. Improving the labeling takes $O(|V|)$ time to find augmenting path and to update the labelling accordingly. We might have to improve the labeling up to times if there is no augmenting path. This makes for a total of $O(|V|^2)$ time. In all, there are $O(|V|)$ iterations each taking $O(|V|)$ work, leading to a total running time of $O(|V|^3)$. This show a great improve to the brute force algorithm, which has a time complexity of $O(|V|!)$.
	
\section{conclusion}
Due to the inefficiency of the traditional manual method and the brute force method, we propose UHA to efficiently solve the assignment problem. Compared with brute force method, we improve the time complexity from $O(N!)$ to $O(N^3)$.
\par
When using this algorithm to solve assignment problem, we start with some Matching $M$, a valid labeling $l$, where $l$ is defined as the feasible labelling. Then looking for an augmenting path in $M$ until a best matching is found, if an augmenting path does not exist, improve the labeling and then go back to first step.
\par
However, our method is not suitable for the assignment problems which the tasks have specific orders. In the future, we plan to optimize our algorithm to solve this problem.

	
	%

	\ifCLASSOPTIONcaptionsoff
	\newpage
	\fi

\end{document}